\documentclass[%
 reprint,
 preprintnumbers,
 amsmath,amssymb,
 aps,
 prl
]{revtex4-2}

\usepackage{graphicx}
\usepackage{bm}
\usepackage{hyperref}

\usepackage{mathtools}
\usepackage{xspace}
\newcommand{\nnlojet}{NNLO\protect\scalebox{0.8}{JET}\xspace}
\newcommand{\rapidix}{RapidiX\xspace}

\DeclareRobustCommand{\ensuremathrm}[1]{\ensuremath{\mathrm{#1}}\xspace}

\DeclareRobustCommand{\ensuremathcal}[1]{\ensuremath{\mathcal{#1}}\xspace}
\DeclareRobustCommand{\rd}{\ensuremathrm{d}} 
\DeclareRobustCommand{\rT}{\ensuremathrm{T}} 

\DeclareMathOperator{\arccosh}{arccosh}
\DeclareRobustCommand{\cO}{\ensuremathcal{O}\xspace}
\DeclareRobustCommand{\LO}{\text{LO}\xspace}
\DeclareRobustCommand{\NLO}{\text{NLO}\xspace}
\DeclareRobustCommand{\NNLO}{\text{NNLO}\xspace}
\DeclareRobustCommand{\N}[1]{\ensuremath{\text{N}^{#1}}} 
\DeclareRobustCommand{\jet}{\text{jet}\xspace}

\DeclareRobustCommand{\mur}{\ensuremath{\mu_\mathrm{R}}\xspace}
\DeclareRobustCommand{\muR}{\mur}
\DeclareRobustCommand{\muf}{\ensuremath{\mu_\mathrm{F}}\xspace}
\DeclareRobustCommand{\muF}{\muf}
\DeclareRobustCommand{\PH}{{\ensuremathrm{H}}\xspace}

\DeclareRobustCommand{\Pgg}{{\ensuremathrm{\gamma}}\xspace}

\DeclareRobustCommand{\Pt}{{\ensuremathrm{t}}\xspace}

\DeclareRobustCommand{\GeV}{\ensuremathrm{GeV}\xspace}
\DeclareRobustCommand{\TeV}{\ensuremathrm{TeV}\xspace}

\DeclareRobustCommand{\FS}{\ensuremath{F}\xspace}

\begin{document}

\preprint{CERN-TH-2021-021, IPPP/20/80, KA-TP-25-2021, NIKHEF 2021-005, SLAC-PUB-17588, ZU-TH 07/21}

\title{Fully Differential Higgs Boson Production to Third Order in QCD}

\author{%
  X. Chen$^{a,b,c}$,
  T. Gehrmann$^a$,
  E.W.N. Glover$^d$,
  A. Huss$^e$,
  B. Mistlberger$^f$,
  A. Pelloni$^g$
}
\affiliation{%
  $^a$\mbox{Department of Physics, Universit\"at Z\"urich, Winterthurerstrasse 190, CH-8057 Z\"urich, Switzerland} \\
  $^b$\mbox{Institute for Theoretical Physics, Karlsruhe Institute of Technology, 76131 Karlsruhe, Germany} \\
  $^c$\mbox{Institute for Astroparticle Physics, Karlsruhe Institute of Technology, 76344 Eggenstein-Leopoldshafen, Germany} \\
  $^d$\mbox{Institute for Particle Physics Phenomenology, University of Durham, Durham DH1 3LE, UK} \\
  $^e$\mbox{Theoretical Physics Department, CERN, 1211 Geneva 23, Switzerland} \\
  $^f$\mbox{SLAC National Accelerator Laboratory, Stanford University, Stanford, CA 94039, USA} \\
  $^g$\mbox{Nikhef Theory Group, Science Park 105, 1098 XG Amsterdam, The Netherlands} \\
}


\begin{abstract}
  We present the first fully differential predictions for the production cross section of a Higgs boson via the gluon fusion mechanism at next-to-next-to-next-to-leading order (\N3\LO) in QCD perturbation theory.
  Differential distributions are shown for the two-photon final state produced by the decay of the Higgs boson for a realistic set of fiducial cuts.
  The \N3\LO corrections exhibit complex features and are in part larger than the inclusive \N3\LO corrections to the production cross section.
  Overall, we observe that the inclusion of the \N3\LO QCD corrections significantly reduces the perturbative uncertainties and leads to a stabilisation of the perturbative expansion.
\end{abstract}

\maketitle


Since the discovery of the Higgs boson by the ATLAS~\cite{Aad:2012tfa} and CMS~\cite{Chatrchyan2012} collaborations at the Large Hadron Collider (LHC), uncovering and understanding the nature of this scalar particle has been a critical objective of the particle physics community.
For the first time, we have the opportunity to directly study the mechanism of electroweak symmetry breaking and ascertain if the properties of this new particle align with our expectations based on the Standard Model (SM) of particle physics.
The ever-increasing amount of data collected at the LHC allows us to probe intricate features of Higgs boson events in great detail (see for example Refs.~\cite{ATLAS:2020qdt,CMS:2020gsy,Sirunyan:2018koj,Aad:2019mbh,Aaboud:2018ezd,Sirunyan:2018sgc}).
Studying the differential distributions of the decay products of the Higgs boson enhances our capabilities to discern the effects of new physics from expectations based on the SM.

To address the fundamental nature of the Higgs boson and to measure its properties, it is of paramount importance to understand theoretically the features of its production and decay
to a degree that rivals or surpasses the precision achieved by the experimental measurements.
To predict the outcome of scattering events at the LHC we use perturbative quantum field theory to relate our fundamental understanding of nature to realistic LHC observables.
At the LHC, the gluon fusion production mechanism (ggF) is responsible for $\sim 90\%$ of all produced Higgs bosons, and making robust and reliable predictions for this contribution is of particular importance.
This mechanism facilitates the production of the Higgs boson via a virtual top quark loop that is formed out of two fusing gluons extracted from the scattering protons.
The hierarchy between the masses of the top quark and the Higgs boson mass allows for efficient calculations by integrating out the degrees of freedom of the top quark~\cite{Inami1983,Shifman1978,Spiridonov:1988md,Wilczek1977}.
This induces a direct coupling of the Higgs boson to gluons via an effective operator~\cite{Chetyrkin:1997un,Schroder:2005hy,Chetyrkin:2005ia,Kramer:1996iq}
that retains an imprint of the top quark through a logarithmic dependence on its mass $m_\Pt$.
Large perturbative corrections due to quantum chromodynamics (QCD) and incredible experimental progress have created an urgent demand to determine this production cross section to a very high perturbative order.
The corresponding inclusive cross section---simply the answer to the question \emph{``How many Higgs bosons are produced in proton collisions?"}---is today known to next-to-next-to-next-to leading order (\N3\LO) in QCD perturbation theory~\cite{Anastasiou:2015ema,Mistlberger:2018etf,Anastasiou:2016cez}. Recently, the \N3\LO predictions for the inclusive rapidity distribution of the Higgs boson were achieved~\cite{Dulat:2018bfe,Cieri:2018oms}.

Predictions for fiducial cross sections that include realistic selection cuts on the final-state decay products of the Higgs boson enable us to compare theoretical predictions directly to experimental observations, circumventing any extrapolation that would otherwise introduce an additional source of uncertainty.
Moreover, in the light of rapidly accumulating data and the upcoming high luminosity phase of the LHC, such differential comparisons present a unique window into the properties of the Higgs boson.
The fast pace of theoretical developments in recent years has produced several approaches~\cite{%
  GehrmannDeRidder:2005cm,*Currie:2013vh,
  Catani:2007vq,
  Czakon:2010td,*Czakon:2011ve,*Boughezal:2011jf,
  Cacciari:2015jma,
  Boughezal:2015eha,*Boughezal:2015dva,*Gaunt:2015pea,
  Caola:2017dug
}
for the differential calculation of hadron-collider processes at next-to-next-to leading order (NNLO) in QCD.
These new methods have enabled precise predictions for a plethora of Higgs boson observables and for different exclusive final states~\cite{Chen:2019wxf,Chen:2014gva,Chen:2016zka,Bizon:2018foh,Caola:2019htj,Boughezal:2015dra,Caola:2015wna,Boughezal:2015aha}.

In this letter, we go beyond the current paradigm and present for the first time fully differential predictions for the gluon fusion production cross section at \N3\LO in QCD perturbation theory.
We obtain this result via the efficient combination of the fully differential calculation of a Higgs boson in association with a hadronic jet at NNLO and the analytic result for the Higgs boson rapidity distribution at \N3\LO~\cite{Dulat:2018bfe} via the so-called Projection-to-Born (P2B) method~\cite{Cacciari:2015jma}.
In particular, we present the extension of the P2B method to the production of an arbitrary color-neutral final state in hadron--hadron collisions.
As the culmination of the above, we present realistic predictions for differential fiducial distributions for the two-photon final state of the Higgs boson.

\section{The Projection-to-Born Method for Color-Neutral Final States}

Fully differential predictions at higher orders in perturbation theory require special treatment for the cancellation of infrared singularities that appear at the intermediate stages of the calculation.
The P2B method accomplishes this through a special projection operation that allows matching an inclusive calculation to a differential calculation at one order lower but with an additional real emission.
This method was initially conceived in the calculation of \NNLO corrections to the vector-boson-fusion process~\cite{Cacciari:2015jma}.
Exploiting the specific dynamics of the VBF process, it used the structure-function approach where the cross section is factorized into kinematically independent deep-inelastic scattering (DIS) sub-processes on each of the incoming beams.
The same structure-function approach was subsequently used to obtain inclusive results at \N3\LO for Higgs~\cite{Dreyer:2016oyx} and di-Higgs~\cite{Dreyer:2018qbw} production in VBF.
The P2B method was extended to one order higher to compute differential predictions at \N3\LO for the DIS process~\cite{Currie:2018fgr,*Gehrmann:2018odt} and also the Higgs decay into a pair of bottom quarks~\cite{Mondini:2019gid}.

In the following, we extend the P2B method to the production of a color-neutral final state \(\FS\) in hadron--hadron collisions where the cross section does not factorize.
The cross section, (multi-) differential in the observable(s) \(\cO\), is decomposed in the P2B method according to the master formula
\begin{align}
  \frac{\rd\sigma^{\N{k}\LO}_{\FS}}{\rd\cO}
  &=
  \left(
    \frac{\rd\sigma^{\N{(k-1)}\LO}_{\FS+\jet}}{\rd\cO} -
    \frac{\rd\sigma^{\N{(k-1)}\LO}_{\FS+\jet}}{\rd\widetilde{\cO}}
  \right) +
  \frac{\rd\sigma^{\N{k}\LO}_{\FS}}{\rd\widetilde{\cO}}
  \;.
  \label{eq:p2b_master}
\end{align}
Here, the mapping \(\cO \xrightarrow{\text{P2B}} \widetilde{\cO}\) uniquely assigns a Born-level configuration to any final state that contains an arbitrary number of accompanying emissions.
The last term in Eq.~\eqref{eq:p2b_master} corresponds to the inclusive calculation of the process that only retains the differential information with respect to the Born variable(s) denoted \(\widetilde{\cO}\).
As can be seen in Eq.~\eqref{eq:p2b_master}, the P2B subtraction scheme has the unique feature that the local un-integrated subtraction term is given by the full, un-integrated real radiation matrix elements themselves.
As a consequence, phase space singularities associated with fully unresolved configurations are cancelled identically, which results in a particularly stable numeric evaluation.

To define the projection, we consider the real-emission phase space with \(n\) additional parton emissions
\begin{align}
  \Phi_{\FS+n}:\quad p_a + p_b \to p_\FS + k_1 + \ldots + k_n \;,
\end{align}
where the \(k_i\) denote the momenta of the emitted partons.
The projected Born phase space \(\widetilde{\Phi}_\FS\) is defined through a rescaling of the incoming parton momenta
\begin{align}
  \tilde{p}_a &= \xi_a p_a , &
  \tilde{p}_b &= \xi_b p_b ,
\end{align}
with \(\tilde{p}_\FS = \tilde{p}_a + \tilde{p}_b\).
In addition to the on-shell constraint, \(\tilde{p}_\FS^2 = p_\FS^2\), we further require the mapping to preserve the rapidity of \(\FS\), \(\tilde{y}_\FS = y_\FS\), which fully determines the projections with
\begin{align}
  \xi_a \xi_b &=
  \frac{p_\FS^2}{2 p_a p_b}
  , &
  \xi_a/\xi_b &=
  \frac{2 p_b p_\FS}{2 p_a p_\FS}
  .
\end{align}
Finally, in case the final state \(\FS\) comprises \(m\) colorless particles, \(p_\FS \to p_1 + \ldots + p_m\), the ``decay products'' transform via the Lorentz transformation
\begin{align}
  \tilde{p}_i^\mu &= \Lambda^\mu{}_\nu(p_\FS,\tilde{p}_\FS) \; p_i^\nu
  \; ,
\end{align}
which
implements the projection onto Born kinematics, $\tilde{p}_\FS^\mu \equiv \Lambda^\mu{}_\nu(p_\FS,\tilde{p}_\FS) \; p_\FS^\nu$, and
is explicitly given as
\begin{align}
  \Lambda^\mu{}_\nu(p_\FS,\tilde{p}_\FS) &=
  g^\mu{}_\nu
  -\frac{2(p_\FS+\tilde{p}_\FS)^\mu(p_\FS+\tilde{p}_\FS)_\nu}{(p_\FS+\tilde{p}_\FS)^2}
  +\frac{2\tilde{p}_\FS^\mu p_{\FS,\nu}}{p_\FS^2}
  \; .
\end{align}

\section{Application to Higgs Production and Phenomenological Results}

\begin{figure}[t!]
\centering
\includegraphics[scale=.85]{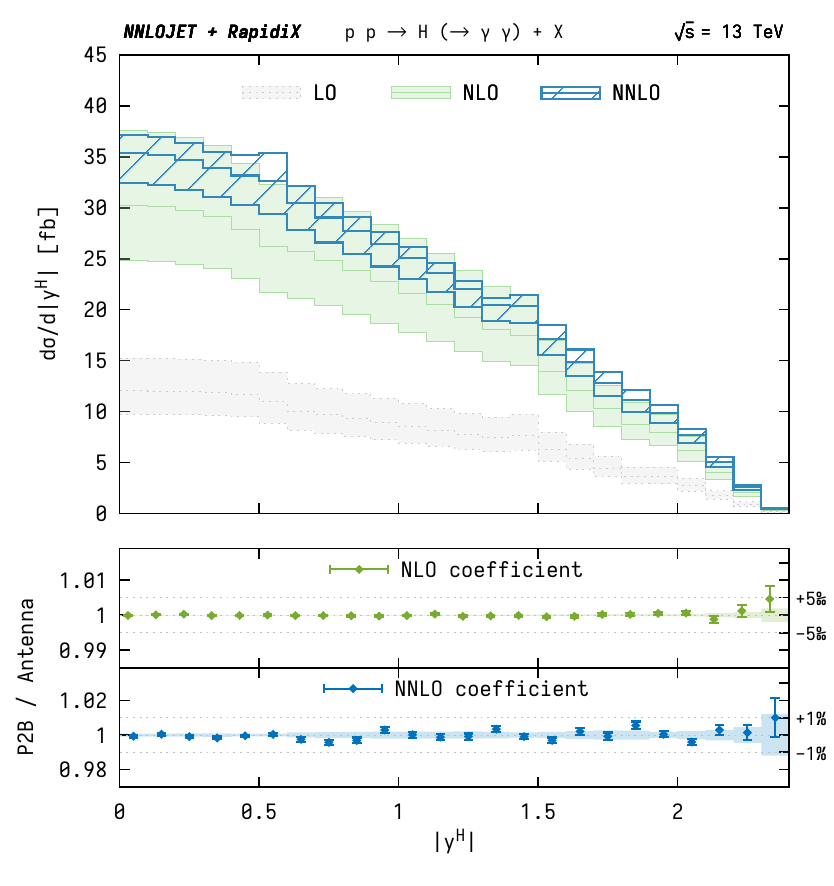}
\caption{\label{fig:validation}%
  Validation of the P2B method against an independent implementation using the antenna-subtraction method up to NNLO.
}
\end{figure}

\begin{figure*}
\hfill
\includegraphics[scale=.85]{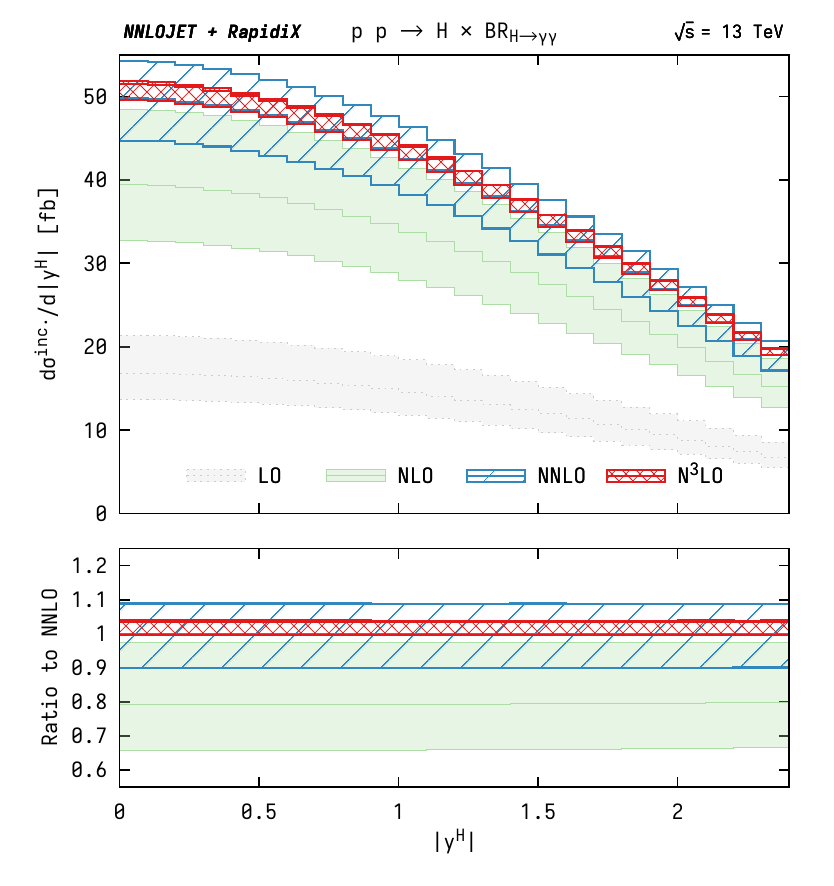}
\hfill
\includegraphics[scale=.85]{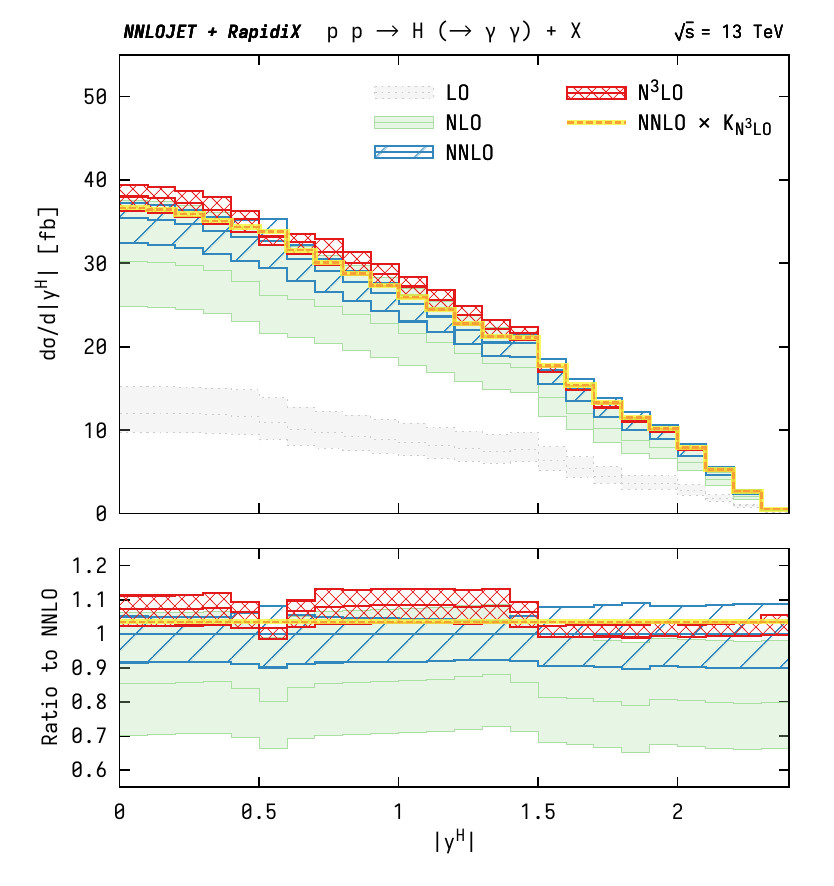}
\hspace*{\fill}
\caption{\label{fig:inc-diff}%
  Comparison between inclusive (left) and fiducial (right) predictions for the rapidity distribution of the Higgs boson up to \N3\LO.
  Predictions are shown at \LO~(grey), \NLO~(green), \NNLO~(blue), \N3\LO~(red), and for the \NNLO prediction re-scaled by the inclusive $K_{\N3\LO}$-factor~(orange).
}
\end{figure*}

We apply the P2B method described in the previous section to Higgs boson production in the gluon-fusion channel at \N3\LO.
The fully differential prediction is assembled according to Eq.~\eqref{eq:p2b_master}, which requires:
\begin{enumerate}
\item
The inclusive calculation at \N3\LO for the Higgs rapidity distribution \(y^\PH\) as computed in Ref.~\cite{Dulat:2018bfe} and implemented in the \rapidix library.
This result is based on techniques developed in Refs.~\cite{Dulat:2017brz,Dulat:2017prg} and is given by analytic formulae for the partonic rapidity distribution computed by means of a threshold expansion. We supplement this result by exploiting the fact that the Higgs boson decays isotropically in its rest frame to generate the inclusive \N3\LO calculation differential in the Higgs boson decay products.
\item
The fully differential \NNLO calculation for the \(\PH+\jet\) process.
This has been computed in Ref.~\cite{Chen:2014gva,*Chen:2016zka} using the antenna subtraction method~\cite{GehrmannDeRidder:2005cm,*Daleo:2006xa,*Currie:2013vh} and is available within the parton-level Monte Carlo generator \nnlojet.
\end{enumerate}
We have implemented the P2B method for color-neutral final states within the \nnlojet framework together with an interface to the \rapidix library to access the inclusive part of the calculation.

For our phenomenological results, we restrict ourselves to the decay of the Higgs boson into a pair of photons and closely follow the corresponding 13~\TeV ATLAS measurement~\cite{Aaboud:2018xdt} with the following fiducial cuts
\begin{align}
  p_\rT^{\Pgg_1} &> 0.35 \times m_{\Pgg\Pgg} , &
  p_\rT^{\Pgg_2} &> 0.25 \times m_{\Pgg\Pgg} , \\
  \lvert\eta^\Pgg\rvert &< \mathrlap{2.37 \text{\; excluding\; \(1.37 < \lvert\eta^\Pgg\rvert < 1.52\)} ,} \nonumber
\end{align}
where \(\Pgg_1\) and \(\Pgg_2\) respectively denote the leading and sub-leading photon with \(m_{\Pgg\Pgg} \equiv M_\PH = 125~\GeV\) the invariant mass of the photon-pair system.
For each photon, an additional isolation requirement is imposed where the scalar sum of partons with \(p_\rT>1~\GeV\) within a cone of \(\Delta R = 0.2\) around the photon has to be less than \(5\%\) of the \(p_\rT\) of the photon.
Note that this setup induces a highly non-trivial interplay between the final-state photons and QCD emissions, requiring a fully differential description of the process.
Throughout this letter, we work in the narrow width approximation to combine the production and decay of the Higgs boson.
To derive numerical predictions we use \verb|PDF4LHC15_nnlo_100|~\cite{Butterworth:2015oua} parton distribution functions and choose the value of the top quark mass in the modified minimal subtraction scheme to be $m_\Pt(m_\Pt)=162.7~\GeV$.

Figure~\ref{fig:validation} compares predictions for the fiducial rapidity distribution of the Higgs boson \(y^\PH\) based on two different methods.
This comparison serves as the validation of the P2B implementation up to \NNLO against an independent calculation based on the antenna subtraction method.
The lower panels in Fig.~\ref{fig:validation} show the ratio of the two calculations, where the filled band and the error bars correspond to the uncertainty estimates of the Monte Carlo integration of the antenna- and P2B-prediction, respectively.
The ratios shown in the bottom two panels reveal agreement within numerical uncertainties between the two calculations at the per-mille and sub-per-cent level for the \emph{coefficients} at \NLO and \NNLO, respectively.

\begin{figure*}
\hfill
\includegraphics[scale=.85]{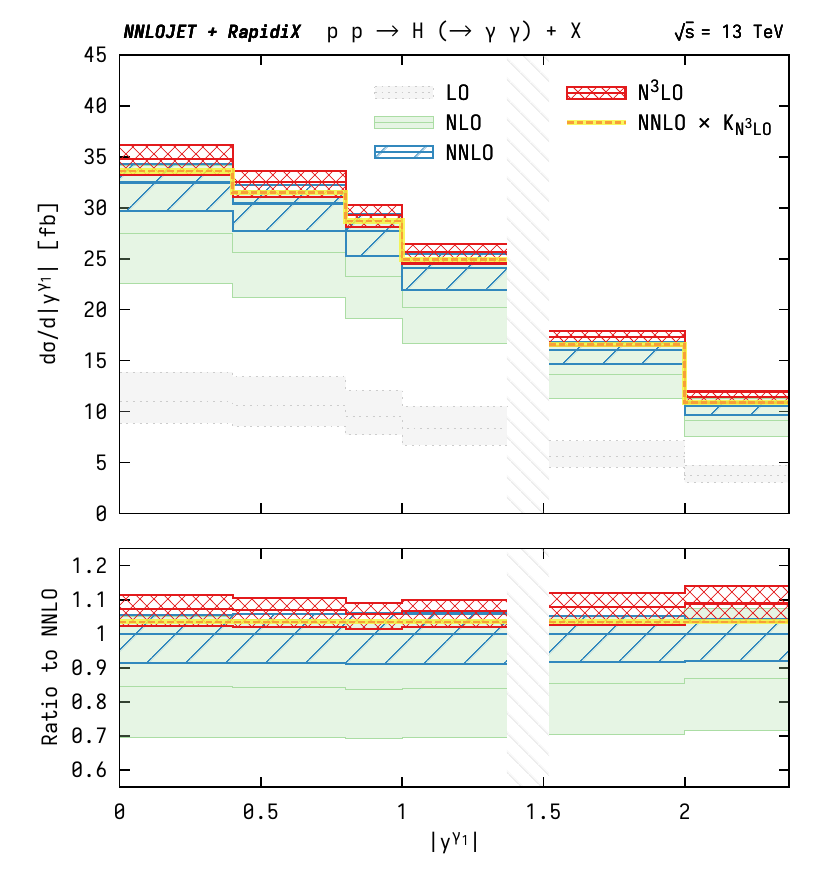}
\hfill
\includegraphics[scale=.85]{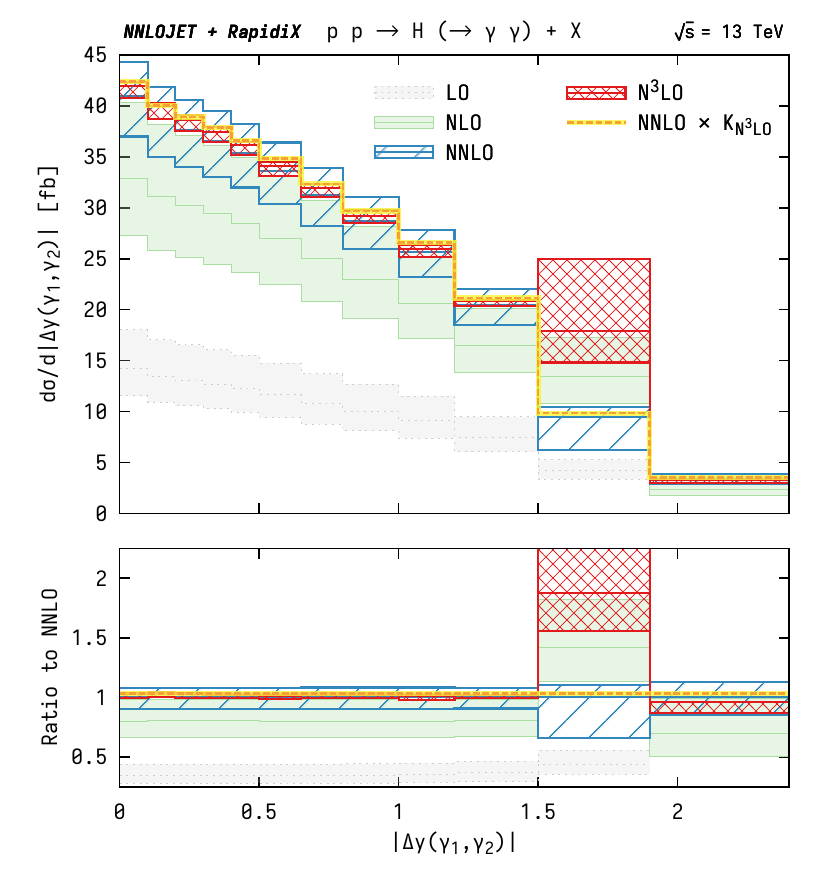}
\hspace*{\fill}
\caption{\label{fig:n3lo}%
  Differential predictions for the rapidity of the leading photon (left) and the absolute value of the difference of the rapidities of the two photons (right).
  Predictions are shown at \LO~(grey), \NLO~(green), \NNLO~(blue), \N3\LO~(red), and for the \NNLO prediction re-scaled by the inclusive $K_{\N3\LO}$-factor~(orange).
}
\end{figure*}

Figure~\ref{fig:inc-diff} compares the inclusive rapidity distribution of the Higgs boson to the fiducial rapidity distribution of the di-photon pair.
It was already noted in Refs.~\cite{Dulat:2018bfe,Cieri:2018oms} that the \N3\LO correction to the inclusive rapidity distribution is remarkably uniform and is well approximated by rescaling the inclusive NNLO rapidity distribution with the inclusive K-factor,
\begin{align}
  K_{\N3\LO} \equiv \sigma_\text{incl.}^{\N3\LO}/\sigma_\text{incl.}^{\NNLO}.
\end{align}
Throughout this letter, we estimate the uncertainty of the truncation of the perturbative series by independently varying the factorization and renormalization scale around their central value $\muF^{\text{cent.}}=\muR^{\text{cent.}}=M_\PH/2$ by factors of \((\tfrac{1}{2},\,2)\) with the restriction $\tfrac{1}{2} \leq \muF/\muR \leq 2$.
The resulting uncertainty estimates for the inclusive predictions are uniform throughout the entire range of the distribution and nearly identical to the uncertainty estimates of $K_{\N3\LO}$.
The right-hand side of Fig.~\ref{fig:inc-diff} shows the fiducial rapidity distribution of the two-photon pair subject to all final-state selection cuts.
The corresponding distribution obtained by re-scaling the NNLO distribution with the inclusive factor $K_{\N3\LO}$ is given by the orange, dashed line.
We observe that this naive treatment of fiducial \N3\LO corrections cannot capture all features of the full result, which are induced by the non-trivial fiducial cuts.
In particular, we observe that in the central region of the rapidity distribution \N3\LO corrections are larger than expected from the inclusive \(K\)-factor.
Furthermore, the obtained estimate of uncertainties due to missing higher-order corrections are slightly larger than in the inclusive case.
This can however be attributed to the fact that the inclusive predictions exhibit a very asymmetric scale variation, potentially underestimating uncertainties.
Nevertheless, we observe that \N3\LO corrections lead to a stabilization of the perturbative expansion and are compatible with NNLO predictions.
Finally, the newly obtained corrections lead to a significant reduction in perturbative corrections.

Figure~\ref{fig:n3lo} shows distributions of the genuine photon final states of the Higgs boson production cross section.
On the left, we show the rapidity distribution of the photon with the leading transverse momentum.
Similar to the fiducial rapidity distribution of the Higgs boson, we observe here that genuine \N3\LO predictions are larger than expected from the inclusive \(K\)-factor, as indicated by the dashed line.
Nevertheless, scale variation bands of \NNLO and \N3\LO predictions overlap and we see that the inclusion of \N3\LO corrections leads to a reduction of the scale dependence.
The right side of Fig.~\ref{fig:n3lo} shows the di-photon cross section as a function of the rapidity difference of the two photons.
This observable displays a perturbative behavior that is very much in unison with the inclusive \(K\)-factor, except for its penultimate bin that exhibits a perturbative instability.
The origin of this instability can be traced back to a linear dependence of the fiducial acceptance on the Higgs transverse momentum (\(p_\rT^{\PH}\))~\cite{Ebert:2019zkb} that induces a sensitivity to very low momentum scales.
As such, the location of the instability is confined to the region close to the kinematic boundaries of the associated process where \(p_\rT^{\PH}\) is very small, also known as so-called ``Sudakov shoulders''~\cite{Catani:1997xc}.
The minimum \(p_\rT^\text{cut}\) on the photons leads to an implicit restriction of the Born-level kinematics, such that vanishing \(p_\rT^{\PH}\) can only be attained below a critical value of \( \lvert \Delta y (\Pgg_1,\Pgg_2) \rvert \).
This boundary is located at \( 2\arccosh\bigl(\tfrac{M_\PH}{2\,p_\rT^\text{cut}}\bigr) \equiv \Delta y_\mathrm{max}\rvert_\LO \approx 1.8 \) and the last bin in Fig.~\ref{fig:n3lo} is thus only populated starting from NLO.
It is easy to verify that this kinematic boundary can only be probed for Higgs rapidities that satisfy
\( \lvert y^\PH \rvert \leq \eta^\Pgg_\mathrm{max} - \tfrac{1}{2} \Delta y_\mathrm{max}\rvert_\LO \approx 1.5 \), which corresponds precisely to the region of \(y^\PH\) where larger higher-order corrections are observed.
Furthermore, we can identify a second region in which the barrel--endcap rejection (\( y^\Pgg_\mathrm{BE} \sim 1.45 \)) crosses over the \( \Delta y \) boundary at
\( \lvert y^\PH \rvert \sim \eta^\Pgg_\mathrm{BE} - \tfrac{1}{2} \Delta y_\mathrm{max}\rvert_\LO \approx 0.55 \),
which aligns precisely with the pronounced dip seen in the Higgs rapidity distribution.
To what extent such a sensitivity can also affect the other observables as well as approaches to avoid them will be left for future studies.

\section{Conclusions and Outlook}

In this letter, we extended the Projection-to-Born method to production cross sections for generic colorless final states at the LHC.
We applied this method and derived for the first time fully differential predictions for a genuine LHC $2\to 1$ production cross section at \N3\LO in QCD perturbation theory.
In particular, we predict the cross section for the production of a Higgs boson and its subsequent decay to final-state photons at \N3\LO.
To achieve this we combine fully differential predictions for the production of a Higgs boson in association with a hadronic jet and the prediction of the inclusive rapidity distribution.

Our result is a significant improvement of the description of some of the most relevant Higgs boson observables.
In particular, we find that \N3\LO corrections for fiducial distributions for the two-photon final state can be non-uniform across the different distributions.
Overall, our newly obtained predictions lead to a reduction in the dependence of the differential cross section on perturbative scales and we find that \N3\LO corrected  predictions are compatible with the \NNLO results within their estimated uncertainties.

In this letter, we investigated specifically the impact of \N3\LO QCD corrections on fiducial cross section predictions.
To perform a direct comparison with LHC observations, these predictions need to be combined with electroweak corrections and effects due to neglected heavy quark masses.
Furthermore, a careful study of sources of uncertainties beyond perturbative corrections is required and we look forward to achieving this in future work.

\emph{Note added}.---During the completion of this work, we became aware of complementary work on the computation of the fiducial cross section for Higgs boson production to third order in QCD employing the \(q_\rT\)-subtraction approach~\cite{Billis:2021ecs}.

\section*{Acknowledgments}

The authors XC, AH, and BM would like to express a special thanks to the Mainz Institute for Theoretical Physics (MITP) for its hospitality and support during the workshop ``High Time for Higher Orders: From Amplitudes to Phenomenology''.
BM was supported by the Department of Energy, Contract DE-AC02-76SF00515.
This work was supported in part by the Swiss National Science Foundation (SNF) under contract 200020-175595.

\bibliography{Hto2p_p2b}

\end{document}